\documentclass[lettersize,journal]{IEEEtran}
\usepackage{amsmath}
\usepackage{amsfonts}
\usepackage{algorithm}
\usepackage{array}
\usepackage[caption=false,font=normalsize,labelfont=sf,textfont=sf]{subfig}
\usepackage{textcomp}
\usepackage{stfloats}
\usepackage{url}
\usepackage{verbatim}
\usepackage{graphicx}
\usepackage{cite}
\usepackage{listings}
\usepackage{graphicx}
\usepackage{booktabs}
\usepackage{multirow}
\usepackage{url}
\usepackage{comment}
\usepackage{microtype}
\usepackage{pifont}
\newcommand{\cmark}{\ding{51}}  % Check mark
\newcommand{\xmark}{\ding{55}}  % Cross mark
\usepackage{xcolor}
\usepackage{tabularx}
\usepackage{makecell}
\usepackage[noend]{algpseudocode}
\usepackage{xcolor}

\definecolor{codegreen}{rgb}{0,0.6,0}
\definecolor{codegray}{rgb}{0.5,0.5,0.5}
\definecolor{codepurple}{rgb}{0.58,0,0.82}
\definecolor{backcolour}{rgb}{0.95,0.95,0.92}
\lstdefinestyle{mystyle}{
  backgroundcolor=\color{backcolour}, commentstyle=\color{codegreen},
  keywordstyle=\color{black},
  numberstyle=\tiny\color{codegray},
  stringstyle=\color{codepurple},
  basicstyle=\ttfamily\footnotesize,
  breakatwhitespace=false,         
  breaklines=true,                 
  captionpos=b,                    
  keepspaces=true,                 
  numbers=left,                    
  numbersep=5pt,                  
  showspaces=false,                
  showstringspaces=false,
  showtabs=false,                  
  tabsize=2
}
\usepackage{algpseudocode}
\begin{document}
\title{RAG Meets Temporal Graphs: Time-Sensitive Modeling and Retrieval for Evolving Knowledge}
\author{Jiale~Han,
	Austin Cheung,
	Yubai Wei,
    Zheng Yu,
    Xusheng Wang,
    Bing Zhu,
	and~Yi~Yang 
	\IEEEcompsocitemizethanks{
		\IEEEcompsocthanksitem 
		Jiale Han, Austin Cheung, and Yi Yang are with the Hong Kong University of Science and Technology, Hong Kong, China 999077. E-mail: jialehan@ust.hk, mycheungaf@connect.ust.hk, imyiyang@ust.hk. 
		\IEEEcompsocthanksitem Yubai Wei is with the University of Turku, Finland FI-20014. E-mail: yubwei@utu.fi.
		\IEEEcompsocthanksitem Zheng Yu, Xusheng Wang, and Bing Zhu are with the HSBC Holdings Plc., Emerging Technology, Innovation, and Ventures, China 200120. E-mail: \{matthew.z.yu, atlas.x.wang, bing1.zhu\}@hsbc.com.} 
} 



\maketitle

\begin{abstract}
Knowledge is inherently time-sensitive and continuously evolves over time. Although current Retrieval-Augmented Generation (RAG) systems enrich LLMs with external knowledge, they largely ignore this temporal nature. This raises two challenges for RAG. First, current RAG methods lack effective time-aware representations. Same facts of different time are difficult to distinguish with vector embeddings or conventional knowledge graphs. Second, most RAG evaluations assume a static corpus, leaving a blind spot regarding update costs and retrieval stability as knowledge evolves. To make RAG time-aware, we propose Temporal GraphRAG (TG-RAG), which models external corpora as a bi-level temporal graph consisting of a temporal knowledge graph with timestamped relations and a hierarchical time graph. Multi-granularity temporal summaries are generated for each time node to capture both key events and broader trends at that time. The design supports incremental updates by extracting new temporal facts from the incoming corpus and merging them into the existing graph. The temporal graph explicitly represents identical facts at different times as distinct edges to avoid ambiguity, and the time hierarchy graph allows only generating reports for new leaf time nodes and their ancestors, ensuring effective and efficient updates. During inference, TG-RAG dynamically retrieves a subgraph within the temporal and semantic scope of the query, enabling precise evidence gathering. Moreover, we introduce ECT-QA, a time-sensitive question-answering dataset featuring both specific and abstract queries, along with a comprehensive evaluation protocol designed to assess incremental update capabilities of RAG systems. Extensive experiments show that TG-RAG significantly outperforms existing baselines, demonstrating the effectiveness of our method in handling temporal knowledge and incremental updates.\footnote{The dataset and code are available at: \url{https://github.com/hanjiale/Temporal-GraphRAG}.}
\end{abstract}


\section{Introduction}
\IEEEPARstart{R}{etrieval}-Augmented Generation (RAG) \cite{DBLP:conf/nips/LewisPPPKGKLYR020, DBLP:journals/corr/abs-2312-10997} has emerged as a powerful paradigm that equips large language models (LLMs) \cite{DBLP:conf/nips/BrownMRSKDNSSAA20, DBLP:journals/corr/abs-2302-13971, guo2025deepseek, DBLP:journals/corr/abs-2505-09388} with external knowledge, which can mitigate LLMs' challenges such as hallucination \cite{DBLP:conf/emnlp/0001PCKW21,DBLP:conf/emnlp/SongWZWCZN24}, limited domain-specific expertise \cite{DBLP:conf/naacl/BhushanNKGPRJ25}, and outdated knowledge \cite{DBLP:conf/acl/OuyangPCYLLL25}. A typical RAG framework consists of three core components: indexing an external corpus into a vector database, retrieving passages that are semantically similar to the input query, and generating an answer conditioned on both the query and the retrieved content. Recent advancements such as GraphRAG \cite{DBLP:journals/corr/abs-2404-16130,11150539} further improve the retrieval process by constructing the original corpus as a knowledge graph. This graph-based representation connects isolated facts through shared entities and relations, enabling multi-hop reasoning and deeper understanding of the corpus.

\begin{figure}[t!]
\centering
\includegraphics[width=1\linewidth]{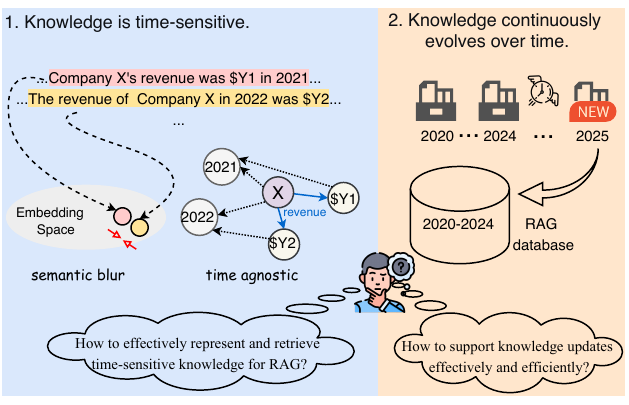}
\caption{The temporal challenges for RAG. (1) Time-aware representation: embeddings for time-sensitive facts are often indistinguishable, and knowledge graphs lack explicit temporal attributes. (2) Incremental updates: corpora evolve continuously, but most evaluations assume one-time indexing, creating an evaluation blind spot on update cost and retrieval stability.
} % Illustration of a time-sensitive question and the limitations of current RAG paradigms.
\label{fig:intro} 
\end{figure}

However, existing RAG frameworks overlook a fundamental dimension of knowledge—\textit{time}. In real-world scenarios, knowledge is inherently time-sensitive and continuously evolves over time \cite{DBLP:conf/nips/ChenWWW21,DBLP:journals/tkde/ZhangZZZWZ24}. For instance, a company's financial reports present varying revenue figures across different fiscal years, and grow over time with the continuous release of new documents. Current RAG systems face significant limitations in  represent and retrieve time-sensitive information. For vector-based RAG, factual statements that differ only in the temporal attributes (e.g., ``Company X's revenue was \$Y\textsubscript{1} in 2021.'' vs. ``Company X's revenue was \$Y\textsubscript{2} in 2022.'') tend to generate highly similar embeddings, rendering them indistinguishable during retrieval \cite{DBLP:conf/sigir/KanhabuaA16, DBLP:journals/corr/abs-2502-21024}. For graph-based methods, knowledge graphs represent relational facts through static triples (subject, relation, object). Such time-agnostic triples lack native support for temporal attributes. For example, the above facts would be converted as two triples  (X, revenue, \$Y\textsubscript{1} ) and (X, revenue, \$Y\textsubscript{2}), with no indication of the time involved.

\IEEEpubidadjcol
Furthermore, real-world corpora evolve continuously, requiring RAG systems to support effective and efficient incremental updates. Although such updates are crucial in real deployments, most studies evaluate RAG systems under a static-corpus setup, where indexing is built once and subsequent evaluation measures retrieval and answer performance on the fixed database. This leaves a critical evaluation blind spot, including the computational costs of updates and performance changes induced by newly ingested information. As illustrated in Figure~\ref{fig:intro}, these observations highlight two fundamental temporal challenges for RAG: (1) how to effectively represent and retrieve time-sensitive knowledge, and (2) how to support knowledge updates at minimal cost while maintaining robust performance.

To address these challenges, we propose Temporal GraphRAG (TG-RAG), a novel framework that models real-world knowledge with a bi-level temporal graph built from the corpus. The lower layer is a temporal knowledge graph whose nodes are entities mentioned in the corpus and whose edges are relations annotated with timestamps. Relations between the same entity pair at different times are preserved as distinct edges, capturing historical evolution. The upper layer organizes all timestamps into a hierarchical time graph. Cross-layer edges connect each time node to the relation edges that are active at that time. For each time node, we maintain a temporal summary that aggregates the facts attached to that node and the summaries of its finer-grained descendants, yielding a corpus-level, time-scoped view of the knowledge. TG-RAG is update-friendly by design. When new documents arrive, we extract timestamped relations and merge them into the existing graph, then generate summaries only for new time nodes and incrementally propagate updates to their ancestor time nodes, avoiding expensive re-summarization from scratch \cite{DBLP:journals/corr/abs-2404-16130}. For retrieval, we use two time-aware strategies: local retrieval extracts fine-grained facts within a specified time window, and global retrieval leverages temporal summaries to capture significant events or broader trends. In this way, TG-RAG provides effective temporal representation and efficient incremental updates, bridging the gap between conventional RAG and the temporal nature of real-world knowledge.

In addition, due to the scarcity of datasets for complex time-sensitive question answering, we contribute ECT-QA, a high-quality dataset designed to evaluate temporal reasoning in RAG systems. ECT-QA is derived from earnings-call transcripts spanning many companies and time periods, reflecting dynamic factual knowledge within evolving corporate contexts. It contains both specific queries which are fact-oriented and abstract queries that focus on trends and summarization. This allows for comprehensive evaluation of both precise retrieval and deep contextual understanding.

To simulate incremental knowledge updates in real world scenarios, we split the corpus into two distinct parts. The base corpus includes documents from 2020 to 2023, while the new corpus contains documents from 2024. Queries are correspondingly categorized into base questions and new questions based on whether their answers require information from the new corpus. We conduct three evaluation scenarios to thoroughly assess system performance: evaluating base queries on the base corpus, base queries on the fully updated corpus, and new queries on the updated corpus. This comprehensive evaluation framework effectively addresses previously existing blind spots in RAG evaluation, providing insights into both computational update costs and system stability after updates.

Comprehensive experiments on the ECT-QA dataset show that TG-RAG substantially outperforms current RAG baselines across all evaluation settings, demonstrating the effectiveness of our approach. In particular, when handling evolving knowledge, our method maintains stable performance on historical queries while achieving strong results on new ones with efficient update costs, which highlights its robustness and practicality in dynamic real-world scenarios. Ablation studies and case analyses reveal that these benefits stem from our temporal knowledge graph modeling, which utilizes timestamped edges to encode fine-grained facts and temporal subgraph retrieval to deliver precise evidence gathering. Furthermore, experimental results on widely-used question answering benchmarks and with different LLM backbones confirm the strong generalization capability of our method.

Our contributions are summarized as follows:
\begin{itemize}
    \item \textbf{Make RAG Time-Aware.} We propose Temporal GraphRAG, a novel framework that models external knowledge as a bi-level temporal graph, explicitly representing fine-grained temporal facts while supporting incremental updates for evolving information.

    \item \textbf{Benchmark and Protocol for Incremental Evaluation.} We release ECT-QA, a high-quality benchmark tailored for time-sensitive question answering, along with a comprehensive evaluation protocol specifically designed to assess incremental update capabilities in RAG systems.

    \item \textbf{Effective Retrieval with Efficient Update Cost.} Extensive experiments demonstrate that our approach maintains consistent performance across knowledge updates while achieving low incremental update cost, underscoring its practicality in real-world evolving corpora.
\end{itemize}

\section{Related Work}
\subsection{Retrieval‑Augmented Generation}
Retrieval-Augmented Generation (RAG) has been proposed to connect LLMs with external knowledge sources. Specifically, external corpora are segmented into chunks \cite{DBLP:journals/corr/abs-2402-05131}, embedded \cite{DBLP:conf/acl/WeiHY25}, and stored in a vector database. When a user query arrives, relevant external data is retrieved and combined with the query as input to the LLM, which then generates an informed answer. Beyond this baseline framework, a variety of methods have been developed to further optimize the RAG pipeline. RQ-RAG \cite{DBLP:journals/corr/abs-2404-00610} refines queries for RAG by introducing query rewriting, decomposition, and disambiguation. Self-RAG \cite{DBLP:conf/iclr/AsaiWWSH24} extends the RAG paradigm by enabling LLMs to adaptively decide when and what to retrieve through self-reflection. LumberChunker \cite{DBLP:conf/emnlp/DuarteMGF0O24} leverages  LLMs to dynamically chunk documents by identifying semantic shift points. While these approaches enhance the retrieval of explicit facts, they often fall short when addressing global queries that require connecting multiple implicit facts spread across documents.

To tackle this limitation, Microsoft's GraphRAG \cite{DBLP:journals/corr/abs-2404-16130} constructs an entity knowledge graph from external documents and generates community-level summaries for clusters of closely related entities, thereby enhancing query-focused summarization. MemoRAG \cite{DBLP:journals/corr/abs-2409-05591} leverages a long-context LLM to build a global memory of the corpus, which produces draft answers and guides retrieval tools to locate useful evidences. LightRAG \cite{DBLP:journals/corr/abs-2410-05779} integrates a graph-based text indexing paradigm with a dual-level retrieval framework, enabling efficient and semantically relevant retrieval to better handle complex queries. HippoRAG \cite{DBLP:conf/nips/GutierrezS0Y024} converts the corpus into a schemaless knowledge graph, allowing information integration across passages to support reasoning-oriented retrieval. And HippoRAG 2 \cite{gutierrez2025from} enhances the original HippoRAG by integrating deeper passage-level context into the Personalized PageRank algorithm. However, few studies explicitly model the temporal attributes of external facts, which limits their effectiveness in domains where knowledge evolves rapidly. To address this, we propose TG-RAG, a novel framework that models the dynamics of knowledge through a bi-level temporal graph, significantly enhancing LLMs' retrieval and reasoning performance in such environments.
\begin{figure*}[t!]
\centering
\includegraphics[width=0.7\linewidth]{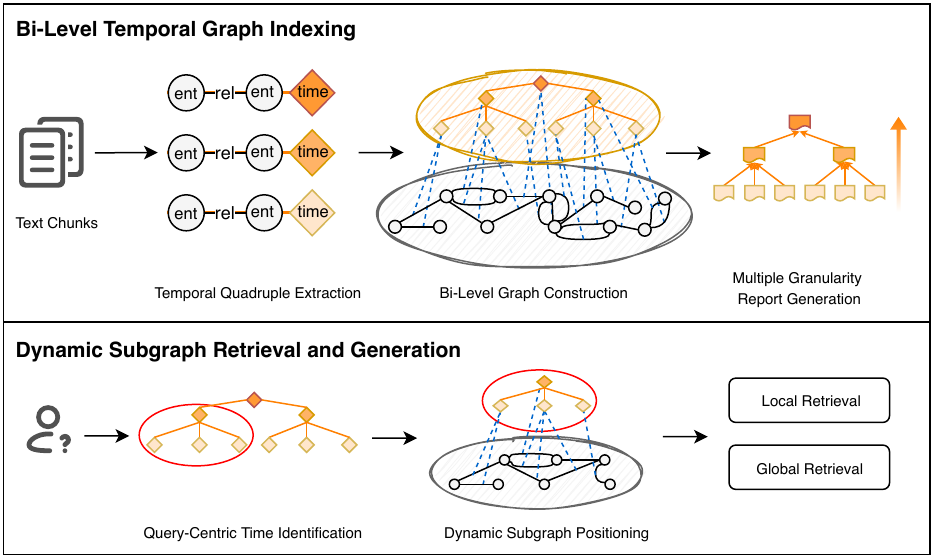}
\caption{The overall framework of Temporal GraphRAG.}
\label{model} 
\end{figure*}

\subsection{Time-Sensitive Question Answering}
Time is an important dimension in our physical world. Lots of facts can evolve with respect to time, including those in finance \cite{DBLP:conf/icaart/PeiZC25}, law \cite{khan2009temporality,stewart2012chronolawgy}, and healthcare \cite{turner2024everything}. Therefore, it is important to consider the time dimension and empower the existing question answering (QA) models to reason over time. Son and Oh \cite{DBLP:conf/emnlp/SonO23} enhance QA models’ temporal understanding through a time-context-dependent span extraction task trained on synthetic temporal data and contrastive time representation learning. Zhu \textit{et al.} \cite{DBLP:conf/emnlp/ZhuYCLLY23} reframe time-sensitive QA as programming, using LLMs to translate natural language into executable code that encodes temporal constraints and selects answers via program execution. Yang \textit{et al.} \cite{DBLP:conf/emnlp/YangLFC24} tackle time-sensitive QA by boosting LLMs’ temporal sensitivity and reasoning with a temporal information-aware embedding and granular contrastive reinforcement learning. Zhang \textit{et al.} \cite{DBLP:journals/corr/abs-2412-15540} point out that current retrievers struggle with temporal reasoning-intensive questions. The authors further propose a training-free modular retrieval framework that decomposes questions into content and temporal constraints, retrieves and summarizes evidence accordingly, and ranks candidates with separate semantic and temporal scores. 

On the dataset side, a few benchmarks have been introduced to systematically probe models’ temporal understanding and reasoning. Chen \textit{et al.} \cite{DBLP:conf/nips/ChenWWW21} build the first time-sensitive QA dataset TimeQA to investigate whether existing models can understand time-sensitive facts. The dataset is constructed by mining time-evolving facts from Wikidata, aligning them to corresponding Wikipedia pages, having crowdworkers verify and calibrate the temporal boundaries, and finally generating question–answer pairs from the annotated time-sensitive facts. Wei \textit{et al.} \cite{DBLP:conf/emnlp/WeiSMYLZZL23} propose a dataset called Multi-Factor Temporal Question Answering (MenatQA) containing multiple time-sensitive factors (scope factor, order factor, counterfactual factor), which can be used to  evaluate the temporal understanding and reasoning capabilities of QA systems. However, most existing datasets contain only simple single-hop time-sensitive queries, highlighting the need for a more challenging benchmark.

\section{Temporal Graph Retrieval-Augmented Generation}
In this section, we present Temporal Graph Retrieval-Augmented Generation (TG-RAG) and begin by formalizing the task. Given a corpus $\mathcal{D}$ and a time-sensitive query $q$, the goal is to retrieve the set of relevant evidences from $\mathcal{D}$ and generate the answer $a$. TG-RAG operates in two stages, a bi-level temporal graph indexing stage that organizes the corpus into a semantically and temporally structured graph, and a dynamic subgraph retrieval and generation stage where time-relevant facts are retrieved to facilitate response generation. The overall architecture of TG-RAG is illustrated in Figure~\ref{model}.

\subsection{Bi-Level Temporal Graph Indexing}
In this stage, we reorganize the corpus into a bi-level temporal graph that supports explicit temporal facts representation and deep time-aware understanding, and enables incremental corpus updates.

\subsubsection{Temporal Quadruple Extraction}
We first segment the raw corpus $\mathcal{D}$ into multiple chunks $\mathcal{C}$ and prompt an LLM to extract temporal quadruples. For each chunk $c \in \mathcal{C}$, the LLM is asked to first recognize timestamp nodes, then identify the non-temporal entities, and detect the temporal relations linking those entities. The output is a set of quadruples $(v_1,\;v_2,\;e,\;\tau)$, 
where $v_1$ and $v_2$ are two entities, $e$ denotes the relation, and $\tau$ is the normalized timestamp. The detailed prompt is presented in the Appendix. %Prompt放在附录中

\subsubsection{Bi-Level Graph Construction} Based on the extracted temporal quadruples, we construct a temporal knowledge graph $\mathcal{G}_K = (\mathcal{V}, \mathcal{E})$, where $\mathcal{V}$ is the set of non-temporal entities and
\(\mathcal{E}=\{(v_1,\;v_2,\;e,\;\tau)\}\) is the set of relation edges annotated with timestamps. Multiple facts between the same entity pair at different times are kept as parallel temporal edges, capturing the evolution of their relationship. Furthermore, all unique timestamps are organized into a time hierarchy graph $\mathcal{G}_T$ that partitions the timeline into nested buckets and connects the parent intervals to their immediate sub-intervals (e.g., \emph{year} $\rightarrow$ \emph{quarter} $\rightarrow$ \emph{month} $\rightarrow$ \emph{day}). Cross-layer edges connect each time node in $\mathcal{G}_T$ to the  edges in $\mathcal{G}_K$ that are active at that time, yielding a bi-level structure that binds factual relations to their temporal scopes. In addition, we precompute and store vector embeddings for entities and relations based on their textual description:
\(\mathbf{e}_v\) for each \(v\in\mathcal{V}\) and \(\mathbf{e}_e\) for each edge in
\(\mathcal{E}\).

\subsubsection{Multi-Granularity Time Report Generation} 
For each time node in $\mathcal{G}_T$ from bottom to top, we generate a report that summarizes the activity within its time window. Specifically, we aggregate all entities and relations in $\mathcal{G}_K$ connected to that time node and the time reports of its child nodes, and prompt an LLM to produce a concise report highlighting noteworthy events, interactions, and trends for that time period.  These summaries propagate recursively upward, forming a multi-granularity hierarchy of reports, progressively offering broader and richer views of the corpus's temporal dynamics.

\subsubsection{Incremental Update of the Temporal Graph}
When new documents arrive, we extract temporal quadruples and merge the resulting timestamped relations into the existing temporal knowledge graph $\mathcal{G}_K$, creating new time nodes in the time hierarchy graph $\mathcal{G}_T$ only when needed. We then generate time reports for the newly created leaf time nodes and incrementally update reports along their ancestor paths in $\mathcal{G}_T$. The reports of unaffected time nodes remain unchanged. Unlike GraphRAG \cite{DBLP:journals/corr/abs-2404-16130} which requires regenerating all summaries upon every update, our approach avoids full recomputation and maintains high efficiency.

\subsection{Dynamic Subgraph Retrieval and Generation}
Given any query $q$, we first identify the query-specific subgraph by semantic and temporal relevance. Then, we design two retrieval modes. Local retrieval ranks entities and time-valid edges in this subgraph and scores their linked chunks to select fine-grained evidence within the query’s temporal scope. Global retrieval selects salient time nodes and their summaries to provide global context. The selected contexts are passed to the generator to produce the final answer.

\subsubsection{Query-Centric Time Identification}
Given a user query $q$, we prompt an LLM to extract every explicit or implicit temporal expression and determine the full set of timestamps required to answer the question. These timestamps are then aligned to the corresponding nodes in the time hierarchy graph. This process yields a time node set $T^q$.

\subsubsection{Dynamic Subgraph Positioning}
We compute the query embedding \(\mathbf{e}_q\) and retrieve the top \(K\) relation edges from the relation set \(\mathcal{E}\), ranked by cosine similarity \(\gamma_{\varepsilon}=\)\(\cos(\mathbf{e}_q,\mathbf{e}_\varepsilon)\) with precomputed relation embeddings. 
The retrieved edges define a query-specific subgraph \(\mathcal{G}_K^q=(\mathcal{V}^q,\mathcal{E}^q)\), which serves as the basis for subsequent local and global retrieval. To ensure temporal relevance, we further filter edges whose timestamps lie within the query’s temporal scope \(T^q\) and gather their neighboring entities, resulting in a temporally focused seed set \(\mathcal{V}^q_t\). 

\subsubsection{Local Retrieval}
We run Personalized PageRank (PPR) on \(\mathcal{G}^q_K\) with the temporally filtered seed set \(\mathcal{V}^q_t\) as the personalization vector, obtaining a relevance score \(s(v)\) for each entity \(v \in \mathcal{V}^q\). For each edge \(\varepsilon=(v_1,\;v_2,\;e,\;\tau)\in\mathcal{E}^q\), we assign $s(\varepsilon) = \mathbf{1}[\tau \in T^q]\,(s(v_1)+s(v_2))$. Here, \(\mathbf{1}[\cdot]\) denotes the indicator function, which returns 1 when the condition holds and 0 otherwise, so edges with \(\tau \notin T^q\) receive zero score. Let \(\mathcal{E}(c)\) be the set of edges extracted from chunk \(c\) originally, we assign scores for each chunk $c$: $s(c) = w(c)\sum_{\varepsilon\in\mathcal{E}^q} s(\varepsilon)$, where $w(c) = \prod_{\varepsilon\in\mathcal{E}(c)}\bigl(1+\gamma_{\varepsilon}\bigr)$ aggregates the relevance scores of the query and all relation edges to capture the overall semantic importance of all relation edges in chunk \(c\). We select the chunks in descending order of \(s(c)\) until the token count reaches the predefined context window $L_{\text{ctx}}$ to form the context \(\mathcal{C}_q\), and feed it together with the query \(q\) to the LLM to generate the final answer \(a\).  The whole procedure is detailed in Algorithm~\ref{alg:dynamic-subgraph-retrieval-local}.
\begin{algorithm}
\caption{Local Retrieval}
\label{alg:dynamic-subgraph-retrieval-local}
\begin{algorithmic}[1]
\Require Query $q$; subgraph $\mathcal{G}_K^q=(\mathcal{V}^q,\mathcal{E}^q)$; temporal seed entities $\mathcal{V}_t^q$; time nodes $T^q$; chunks $\mathcal{C}$; context window $L_{\text{ctx}}$; LLM for QA $\mathcal{M}$
\Ensure answer $a$
\ForAll{$v \in \mathcal{V}^q$} \Comment{entity scores}
\State $s(v) \gets \textsc{PPR}\big(\mathcal{G}_K^q,\ \mathcal{V}_t^q\big)$ \hfill (1)\label{eq:entity_score}
\EndFor
\ForAll{$\varepsilon=(v_1,v_2, r, \tau)\in\mathcal{E}^q$} \Comment{edge scores}
  \State $s(\varepsilon) \gets \mathbf{1}[\tau \in T^q]\,(s(v_1)+s(v_2))$ \hfill (2)\label{eq:relation_score}
\EndFor
\ForAll{$c\in \mathcal{C}$} \Comment{chunk scores}
    \State $\mathcal{E}(c) \gets \{\,\varepsilon \in \mathcal{E}^q \mid \varepsilon \text{ extracted from } c\,\}$
    \State $s(c) \gets \prod_{\varepsilon\in\mathcal{E}(c)}\bigl(1+\gamma_{\varepsilon}\bigr)\cdot\sum_{\varepsilon \in \mathcal{E}^q} s(\varepsilon)$ \hfill (3)\label{eq:chunk_score}
\EndFor
\State $\mathcal{C}^\downarrow \gets \textsc{SortByScoreDesc}(\mathcal{C}, s)$
\State $\mathcal{C}_q \gets \textsc{GreedyPack}(\mathcal{C}^\downarrow, L_{\text{ctx}})$
\State $a \gets \mathcal{M}(\mathcal{C}_q, q)$
\State \Return $a$
\end{algorithmic}
\end{algorithm}

\begin{table*}[t]
\caption{Comparison of ECT-QA with existing question answering datasets used for RAG.\label{tab:data_comparison}}
\centering
\begin{tabular}{lcccccccc}
\toprule 
\textbf{Dataset} & \textbf{Source} & \textbf{Creation} & \textbf{time coverage} & \textbf{multi-hop} & \textbf{time-sensitive}  & \textbf{abstract}\\
NaturalQuestions \cite{DBLP:journals/tacl/KwiatkowskiPRCP19} & Wikipedia & real Google queries   & before 2018  & \xmark & \xmark & \xmark \\
HotPotQA \cite{DBLP:conf/emnlp/Yang0ZBCSM18} & Wikipedia & crowdsourcing  & before 2017 & \cmark & \xmark  & \xmark\\
2WikiMultiHopQA \cite{DBLP:conf/coling/HoNSA20}  & Wikipedia\&Wikidata & logical rules & before 2010 & \cmark & \xmark  & \xmark \\
MultiHop-RAG \cite{DBLP:journals/corr/abs-2401-15391} & News Articles & LLM synthesis &  Sep-Dec, 2023  & \cmark & Partial  & \xmark \\
TimeQA \cite{DBLP:conf/nips/ChenWWW21} & Wikipedia\&Wikidata & crowdsourcing &  before 2021 & \xmark & \cmark  & \xmark \\
MenatQA \cite{DBLP:conf/emnlp/WeiSMYLZZL23} & Wikipedia\&Wikidata & crowdsourcing &  before 2021  & \xmark & \cmark & \xmark \\
UltraDomain \cite{DBLP:journals/corr/abs-2409-05591} & college textbooks  & LLM synthesis  & Unspecified  &  \xmark  & \xmark    & \cmark \\
\midrule
\multirow{2}*{\textbf{ECT-QA} (ours)} & Earnings Call  & LLM synthesis & \multirow{2}*{2020-2024} 
 & \multirow{2}*{\cmark} & \multirow{2}*{\cmark} & \multirow{2}*{\cmark}  \\
& Transcripts &human review & & & & \\
\bottomrule
\end{tabular}
\end{table*}

\subsubsection{Global Retrieval}
We start by collecting evidences for answering the query. Evidences are formed by: (i) chunks with the highest scores $s(c)$ as detailed in the local retrieval workflow, and (ii) time reports of each node in $T^q$. 
Then, each piece of evidence is processed independently by an LLM to extract a set of atomic points $P_e$, where each point $p\in P_e$ is a tuple: $p=(\text{description}, \text{score}, \text{confidence})$. In here, description is a statement containing the key information, score reflects the point's perceived importance to the query, and the confidence is the LLM's self-assessed certainty in the point's accuracy. We then iteratively remove lowest confidence points until the input would fit within context window, and order the points by importance score. Finally, an LLM is tasked to synthesize the points into a single, structured answer. Details of the prompts used for each experiment can be found in the Appendix.

\section{Complex Temporal Question Answering Dataset for RAG}
To evaluate time-sensitive QA, we construct ECT-QA, a benchmark of specific and abstract temporal questions derived from Earnings Call Transcripts. The dataset is curated through LLM-assisted synthesis and human review to ensure high factual quality. As shown in Table~\ref{tab:data_comparison}, ECT-QA fills a key gap in existing RAG benchmarks by addressing time-sensitive multi-hop and abstract temporal reasoning, which is a critical yet underexplored dimension in current datasets. 

\subsection{Temporal Question Definition}
Following the definition in Chen \textit{et al.} \cite{DBLP:conf/nips/ChenWWW21}, we consider a question to be time-sensitive if it includes a temporal specifier such as ``in 2023'' or ``before 2024'' and altering this specifier would change the correct answer. We design two types of temporal question. 
\begin{itemize}
    \item Specific multi-hop questions, which are fine-grained and fact-based queries that typically require locating and connecting multiple pieces of facts to arrive at an answer. For example, ``\textit{Which quarter saw the highest deferred revenue growth for Autodesk Inc from Q3 2022 to Q3 2023?}''
    \item Abstract questions, which are query-centric summarization that focus on high-level understanding of the dataset rather than isolated facts. An example is ``\textit{How did energy companies navigate cost pressures and enhance profitability across 2024?}'' 
\end{itemize}

\subsection{Corpus Collection}
Earnings Call Transcripts (ECTs) serve as the corpus for our study, chosen for their rich temporal characteristics. These transcripts record detailed quarterly financial information for each company and are publicly available for listed companies. We crawl ECTs released between 2020 and 2024 from the Motley Fool platform\footnote{\url{https://www.fool.com/earnings-call-transcripts/}} and retain the companies with transcripts for every quarter in this five-year window. For each transcript, we only keep the ``prepared remarks'' section and discard the rest. The final corpus contains 480 ECTs from 24 companies in 5 different sectors, with a total of 1.58 million tokens. 

\subsection{Specific Question-Answer Synthesis}
We generate a diverse and high-quality temporal multi-hop question-answer dataset with golden evidence through the below pipeline. 
\subsubsection{Temporal Event Extraction and Alignment}
We begin by prompting GPT-4o-mini\footnote{\url{https://platform.openai.com/docs/models/gpt-4o-mini}} to extract temporal events from each document, represented as keyword–timestamp pairs with associated evidence sentences. To align semantically similar but syntactically different keywords, we encode them using SentenceTransformer\footnote{\url{https://sbert.net/}} and apply HDBSCAN clustering \cite{DBLP:conf/icdm/McInnesH17}, enabling alignment and cross-document linking for constructing multi-hop questions.

\subsubsection{Multi-Hop Question Generation by Temporal and Reasoning Types}
To construct diverse specific queries, we categorize question generation based on both temporal scope and reasoning type. Specifically, we group 3–8 evidence sentences referring to the same event entity into three temporal scopes: single-time (within one time point), multi-time (across multiple periods), and relative-time (before/after a time point). These grouped evidences are then fed into GPT-4o-mini to generate either enumeration or comparison questions, along with a direct and factual answer. In addition, we synthesize unanswerable questions that involve time periods, entities, or evidence not present in the corpus. This process yields a rich set of time-sensitive questions requiring different levels of multi-hop reasoning.

\begin{table}[t]
\caption{Statistics and Examples of ECT-QA.\label{tab:data_statistics}}
\centering
\begin{tabularx}{\linewidth}{llX}
\toprule
\textbf{Category} & \textbf{Subcategory} & \textbf{Count} \\
\midrule
\multirow{2}{*}{Corpus} & Documents & 480 \\
& Tokens & 1.58 Million \\
\midrule
\multirow{9}{*}{\makecell{Specific\\Questions}} & Total & 1,005 \\
\cmidrule(lr){2-3}
& \textit{Temporal Scope} & \\
& \quad Single-time (in) & 483 \\
& \quad Multi-time (between) & 321 \\
& \quad Relative-time (before/after) & 201 \\ 
\cmidrule(lr){2-3}
& \textit{Reasoning Type} & \\
& \quad Comparison & 282 \\
& \quad Enumeration & 462 \\
& \quad Unanswerable & 261 \\
\midrule
\multirow{5}{*}{Examples} & \multicolumn{2}{p{0.8\linewidth}}{What was EPAM Systems, Inc.'s utilization in each quarter after 2024 Q1? (Relative-time \& Enumeration \& 3 hops)} \\
\cmidrule(lr){2-3}
& \multicolumn{2}{p{0.8\linewidth}}{In which quarter did EPAM Systems Inc. achieve the highest GAAP gross margin between 2021 Q2 and 2022 Q1? (Multi-time \& Comparison \& 4 hops)}\\
\midrule
\multirow{3}{*}{\makecell{Abstract\\Questions}} & Total & 100 \\
\cmidrule(lr){2-3}
& Single-time (in) & 43 \\
& Multi-time (between) & 57 \\
\midrule
\multirow{5}{*}{Examples} & \multicolumn{2}{p{0.8\linewidth}}{How did companies in the information technology sector, such as Baidu and EPAM, navigate macroeconomic challenges and sector-specific headwinds in 2023 Q1?}\\
\cmidrule(lr){2-3}
&\multicolumn{2}{p{0.8\linewidth}}{Why did Skechers U.S.A., Inc. achieve record revenue achievements between 2020 and 2022?} 
 \\
\bottomrule
\end{tabularx}
\end{table}

\subsubsection{Automatic and Manual Quality Assurance}
To ensure quality, we filter generated questions using four LLM-assessed criteria: reasoning type match, temporal specificity, evidence necessity, and evidence sufficiency. Only questions meeting all criteria are retained. Answers are verified for factual correctness and regenerated if needed. A final round of manual review and refinement ensures the overall quality. This pipeline yields a temporal multi-hop QA dataset grounded in golden evidence.

\subsection{Abstract Question Synthesis}

To synthesize realistic and diverse queries, we construct abstract questions by first simulating potential user profiles with information needs, and then generating temporal queries that reflect authentic analytical intents.

\subsubsection{Potential User Simulation}

Inspired by prior work \cite{DBLP:journals/corr/abs-2404-16130}, we simulate 10 potential user personas who might interact with the corpus. Given a description of the corpus, we ask the LLM to generate descriptive user profiles, each associated with a distinct information need and analytical objective.

\subsubsection{User-Guided Abstract Question Generation}

We group multiple ECT documents based on shared metadata, such as time period, company, or sector. On average, each group contains 18.22 documents. To address input length constraints, we first use LLM to generate concise summaries for each individual document. Given a user profile and the corresponding set of document summaries, we encourage LLM to synthesize deep questions that such a user might pose when analyzing the content, with a particular focus on temporal ``how'' and ``why'' questions.

\subsubsection{Automatic and Manual Quality Assurance}

Similar to the above procedure, we first perform an automatic quality check by prompting the LLM to evaluate each question along four dimensions: clarity, temporal grounding, analytical depth, and answerability. Questions that pass this validation are then manually reviewed to ensure quality. Finally, we construct 100 abstract questions. Table~\ref{tab:data_statistics} summarizes the overall dataset statistics and provides illustrative examples.

\section{Experimental Setup}
This section presents the experimental setup, covering the incremental evaluation protocol, baseline methods, evaluation metrics, and implementation details.

\subsection{Incremental RAG Evaluation}\label{sec:incremental_evaluation}
To simulate the dynamic nature of real-world knowledge environments where RAG systems must continuously adapt to evolving information, we design an incremental evaluation protocol by partitioning both the corpus and the query set into two temporal slices. Specifically, for our ECT-QA dataset, we establish a base corpus consisting of all 384 documents from 2020 to 2023, and a new corpus from 2024 containing 96 documents as incremental updates. In parallel, we partition queries according to the temporal scope of the facts required for answering them: base queries, whose answers rely only on facts from 2020 to 2023, and new queries, whose answers require facts spanning 2020 to 2024. In total, the dataset includes 1,105 specific queries with 656 base queries and 349 new queries, and 100 abstract queries containing 72 base abstract queries and 28 new abstract queries. This design allows us to evaluate RAG performance under corpus growth from three perspectives: 
\begin{itemize}
    \item \textbf{Base queries on the base corpus}: evaluate system performance under the base corpus setting.  
    \item \textbf{Base queries on the updated corpus}: assess consistency and robustness of RAG systems when new corpus is injected.  
    \item \textbf{New queries on the updated corpus}: measure adaptability of RAG systems in leveraging newly added knowledge.  
\end{itemize}

Overall, this incremental evaluation captures both the stability of RAG systems on previously seen queries and their ability to handle new queries grounded in evolving knowledge, closely reflecting real-world deployment scenarios.

\subsection{Baselines}
We compare our approach with the following state-of-the-art methods:
\begin{itemize}
    \item LLM-GT, which concatenates the question with its gold evidence passages and feeds them to the LLM to generate the answer. This approximates an oracle upper bound for the given LLM generator.
    
    \item NaiveRAG \cite{DBLP:journals/corr/abs-2312-10997}, a standard RAG baseline which chunk the corpus, encode passages with an embedding model, and store embedding vectors into a vector base. At test time, the query is embedded and top-$k$ passages are retrieved by similarity, and the LLM generates the answer from the retrieved context.
    
    \item QD-RAG \cite{DBLP:journals/corr/abs-2507-00355}, which first decomposes the original question into simpler sub-queries, and then runs retrieval independently for each sub-query. The retrieved paragraphs are aggregated together with the original question to the LLM to produce the final answer. 
    
    \item GraphRAG \cite{DBLP:journals/corr/abs-2404-16130}, a graph-enhanced RAG pipeline that uses an LLM to extract entities and relations from the corpus, clusters nodes into communities, and generates community reports to capture global context for retrieval and generation.
    
     \item LightRAG \cite{DBLP:journals/corr/abs-2410-05779}, integrates a graph-based text indexing paradigm with a dual-level retrieval scheme to capture both low- and high-level signals for efficient retrieval.
     \item HippoRAG2 \cite{gutierrez2025from}, first builds an open KG by extracting triples from passages with an LLM, and then applies personalized PageRank over the KG to retrieve nodes and maps them back to passages for the final LLM answer.
\end{itemize}

\subsection{Evaluation Metrics} 
To assess the performance of specific queries, we design three LLM-based metrics to automatically judge the factual accuracy of model responses. A powerful LLM (specifically, GPT-4o-mini) serves as the judge, performing a fine-grained, element-wise comparison between the model's prediction and the ground-truth answer, with reference to the provided evidence. Specifically, we utilize this automated framework to measure the proportion of: (1) \textbf{Correct} elements, factual claims that are accurately supported by the provided evidence and match the required temporal scope; (2) \textbf{Refusal} elements, instances where the model explicitly acknowledges its inability to answer due to lack of evidence; and (3) \textbf{Incorrect} elements, responses containing wrong, unsupported, or hallucinated information. These three metrics sum to 1 for each query evaluation. A good QA system is expected to achieve high Correct score and low Incorrect score. Crucially, we expect the model to appropriately refuse answering rather than providing incorrect information \cite{Kalai2025WhyLMHallucinate}. This refusal behavior represents a safety-aware approach that is particularly valuable in high-stakes domains. Besides, we adopt two widely used non-LLM metrics, including \textbf{ROUGE-L} and $\mathbf{F}_{\mathbf{1}}$. ROUGE-L\footnote{\url{https://github.com/google-research/google-research/tree/master/rouge}} evaluates the overlap of the longest common subsequence between generated and reference answers, and $\mathrm{F}_{1}$ measures the harmonic mean of precision and recall at the token level.

Following previous studies \cite{DBLP:journals/corr/abs-2404-16130,DBLP:journals/corr/abs-2410-05779}, we conduct an LLM-based multi-dimensional comparison method for evaluating abstract queries. Specifically, given the answers from two different methods, we employ GPT-4o-mini as a judge to evaluate answer pairs across three key dimensions: \textbf{Comprehensiveness}, the depth and detail of information provided in the answer, \textbf{Diversity}, the variety of perspectives and insights offered, and \textbf{Temporal Coverage}, the accuracy and completeness in handling time-related aspects of the query. For each dimension, the LLM judge selects a winner between two answers and provides detailed explanations. Based on these criteria, an \textbf{Overall Winner} is determined to provide a holistic assessment of answer quality. We calculate win rates accordingly, ultimately leading to the final results. All prompts used for evaluation are provided in the Appendix to ensure reproducibility. 

\subsection{Implementation Details} In our experiments, the chunk size is set to 1,200 tokens with an overlap of 100 tokens. For local retrieval, we retrieve the top-$K$=20 candidate relations. To ensure a fair comparison, our approach and all baseline models utilize the Gemini-2.5-flash-lite\footnote{\url{https://cloud.google.com/vertex-ai/generative-ai/docs/models/gemini/2-5-flash-lite}} model for the indexing phase, and the GPT-4o-mini model for the query answering phase, unless otherwise specified. For text embedding, we adopt the text-embedding-3-small\footnote{\url{https://platform.openai.com/docs/models/text-embedding-3-small}} model. The total length of retrieved content is limited to 12,000 tokens for local retrieval and 24,000 tokens for global retrieval, with 10\% of the token budget allocated to chunks and 90\% to time reports.

\section{Results and Discussion}
In this section, we conduct comprehensive experiments to answer the following questions:
\begin{itemize}
    \item How does our method perform under different evaluation settings?
    We compare our approach with baselines on both specific queries and abstract queries across three evaluation settings, which demonstrates the effectiveness and efficiency of our method. Detailed in Section~\ref{exp:Q1}.

    \item Why does our method perform well?
    We conduct ablation studies and graph visualizations to better understand the performance gains of our method, demonstrating the importance of temporal graph modeling and time-aware retrieval. The findings are discussed in Section~\ref{exp:Q2}.

    \item How well does our method generalize across different LLMs and datasets?
    We evaluate the generalization of our approach across different LLMs for question answering, and further extend the analysis to generic QA benchmarks beyond the ECT-QA dataset to verify its applicability. See Section~\ref{exp:Q3}.
\end{itemize}

\begin{table*}[t]
\caption{The performance of specific question answering: base queries on the base corpus.\label{tab:local_result_old}}
\centering
\begin{tabular}{l|cc|ccccc}
\toprule 
\multirow{2}*{Model} & \multicolumn{2}{c|}{Index Token Cost} &  \multicolumn{3}{c}{LLM Metrics} & \multicolumn{2}{c}{Non-LLM Metrics} \\
& Prompt \textdownarrow & Completion \textdownarrow & Correct \textuparrow & Refusal & Incorrect \textdownarrow & ROUGE-L \textuparrow & $\mathrm{F}_1$ \textuparrow  \\ 
\midrule
    LLM-GT & -- & -- & 0.902 & 0.023  & 0.075  & 0.626  & 0.647  \\\midrule
    NaiveRAG &-- &-- & 0.385  & 0.325  & 0.290  & 0.375 & 0.366   \\
    QD-RAG  & -- &-- & 0.380  & 0.329  &  0.291 & 0.369  &  0.359 \\
    GraphRAG \cite{DBLP:journals/corr/abs-2404-16130} & 37.1M & 17.7M  & 0.405 & 0.280 & 0.315  &  0.371   & 0.375 \\
    LightRAG \cite{DBLP:journals/corr/abs-2410-05779} & 17.1M & 9.0M & 0.406  & 0.160  & 0.434 & 0.350 & 0.359  \\
    HippoRAG2 \cite{gutierrez2025from}&\textcolor{white}{0}3.2M &1.1M & 0.410  &0.345  &  0.245 & 0.382  & 0.385  \\
    %\textbf{TG-RAG (Ours)} &\textcolor{white}{0}6.3M &7.1M & 0.4703 & 0.4160 & 0.1137 & 0.4019 & 0.4004 \\
    \textbf{TG-RAG (Ours)} &\textcolor{white}{0}6.3M &7.1M & \textbf{0.599} & 0.191 & \textbf{0.210} & \textbf{0.493}  & \textbf{0.490} \\
\bottomrule
\end{tabular}
\end{table*}

\begin{table*}[t!]
\caption{The performance of specific question answering: base queries and new queries on the updated corpus.\label{tab:local_result_new}}
\centering
\renewcommand\tabcolsep{2.6pt}
\begin{tabular}{l|cc|ccccc|ccccc}
\toprule 
\multirow{3}*{Model}& & &\multicolumn{5}{c|}{Base Queries} & \multicolumn{5}{c}{New Queries}\\
& \multicolumn{2}{c|}{Index Token Cost} & \multicolumn{3}{c}{LLM Metrics} & \multicolumn{2}{c|}{Non-LLM Metrics} &  \multicolumn{3}{c}{LLM Metrics} & \multicolumn{2}{c}{Non-LLM Metrics} \\
& Prompt \textdownarrow & Completion \textdownarrow & Correct \textuparrow & Refusal & Incorrect \textdownarrow & ROUGE-L \textuparrow & $\mathrm{F}_1$ \textuparrow  & Correct \textuparrow & Refusal & Incorrect \textdownarrow & ROUGE-L \textuparrow & $\mathrm{F}_1$ \textuparrow   \\ 
\midrule
    LLM-GT & -- &-- & 0.902 & 0.023  & 0.075  & 0.626  & 0.647 & 0.874 & 0.044 & 0.082 & 0.631 & 0.649  \\\midrule
    NaiveRAG & -- &-- & 0.366  & 0.347  & 0.287  & 0.352 & 0.343 & 0.390  & 0.406 & 0.205  & 0.384 & 0.356  \\
    QD-RAG & -- &-- & 0.362  & 0.374 & 0.264  & 0.354   & 0.344 & 0.407 &  0.355 & 0.238 & 0.413 & 0.386 \\
    GraphRAG & 30.0M &7.8M & 0.380 & 0.327 &0.293  & 0.370 & 0.369 & 0.398 & 0.368 & 0.234  & 0.392 & 0.389   \\
    LightRAG  & 4.6M & 2.6M & 0.386  & 0.203   & 0.412   & 0.345   &  0.355  & 0.382  &  0.229 & 0.389 & 0.354 & 0.348   \\
    HippoRAG2  & 0.8M & 0.3M & 0.399  & 0.347  & 0.253  &0.376    & 0.372   &  0.372  & 0.492 & 0.136  &0.367  & 0.354   \\
    %\textbf{TG-RAG (Ours)} & 1.6M & 2.0M & 0.4536 & 0.4403 &  0.1061 & 0.3893  & 0.3922 & 0.4739 & 0.4174  & 0.1087  & 0.4351 & 0.4181 \\
    \textbf{TG-RAG (Ours)} & 1.6M & 2.0M & \textbf{0.587} & 0.193 &  \textbf{0.220} & \textbf{0.483}  & \textbf{0.475} & \textbf{0.617} & 0.216 & \textbf{0.167}  & \textbf{0.524} & \textbf{0.487} \\
\bottomrule
\end{tabular}
\end{table*}

\begin{table}[t]
\caption{The performance of abstract question answering. \textit{Abbreviations:} Comp.=Comprehensiveness, Div.=Diversity, Temp. =Temporal Coverage.\label{tab:global_result}}
\centering
\renewcommand\tabcolsep{2.8pt}
\begin{tabular}{l|cc|cc|cc}
\toprule
\textbf{Model} & GraphRAG & \textbf{Ours} & HippoRAG & \textbf{Ours} & LightRAG & \textbf{Ours} \\
\midrule
\multicolumn{7}{c}{\textit{(1) Base queries on the base corpus}} \\
\midrule
Comp.      & 0.167& 0.833& 0.167 & 0.833 & 0.014& 0.986    \\
Div.             & 0.486&0.514 & 0.431 & 0.569 & 0.097& 0.903    \\
Temp. Cov.     & 0.111 &0.889 & 0.236 & 0.764 &0.014 & 0.986    \\
Overall               & 0.167& 0.833& 0.181 & 0.819 & 0.014& 0.986    \\
\midrule
\multicolumn{7}{c}{\textit{(2) Base queries on the updated corpus}} \\
\midrule
Comp.      & 0.222 & 0.778& 0.167& 0.833& 0.028& 0.972    \\
Div.             & 0.472 &0.528 &0.375 &0.625 & 0.042&0.958     \\
Temp. Cov.     & 0.111 &0.889 &0.264 &0.736 & 0.069&0.972     \\
Overall               &  0.236& 0.764& 0.181& 0.819& 0.028&0.972     \\
\midrule
\multicolumn{7}{c}{\textit{(3) New queries on the updated corpus}} \\
\midrule
Comp.      & 0.107 &0.893 & 0.214&0.786 & 0.000&1.000     \\
Div.             &  0.357& 0.643&0.536 &0.464 & 0.071&0.929     \\
Temp. Cov.     &  0.000& 1.000&0.321 &0.679 &0.107 &0.893     \\
Overall               &  0.107& 0.893& 0.250&0.750 & 0.000&1.000     \\
\bottomrule
\end{tabular}
\end{table}

\subsection{How does our method perform?} \label{exp:Q1}
This sections present and analyze the main results on both specific and abstract question answering tasks, demonstrating our method's superior performance across different question complexities.
\subsubsection{Results of Specific Question Answering} 
Following the evaluation settings introduced in Section~\ref{sec:incremental_evaluation}, we first evaluate the performance of different methods on the base queries using the base corpus (2020–2023). Table~\ref{tab:local_result_old} summarizes the performance of all methods based on LLM-based factual accuracy metrics (Correct, Refusal, and Incorrect), traditional lexical overlap metrics (ROUGE-L and $\mathrm{F}_1$), as well as indexing cost measured by the number of prompt and completion tokens consumed during LLM-based indexing. We observe that LLM-GT achieves very high performance across all metrics, with a Correct ratio of 0.902 and an Incorrect ratio of 0.075. This demonstrates the high factual quality of the ECT-QA dataset, confirming its suitability for reliable evaluation. Compared to the other baselines, our method achieves the highest Correct score of 0.599 and the lowest Incorrect score of 0.210, indicating superior factual accuracy. In terms of cost efficiency during indexing phase, our method maintains competitive token cost with 6.3 million prompt tokens and 7.1 million completion tokens, which are significantly lower than GraphRAG and LightRAG. It is worth noting that HippoRAG2's low indexing cost comes at the expense of semantic richness, as it constrains the LLM to output only entity and relation names, omitting detailed descriptions and high-level summaries. In contrast, our method achieves a more balanced trade-off, which avoids exhaustive graph construction while achieving high retrieval performance.

Next, we inject the new corpus from 2024 into the existing database to simulate the real-world scenario where RAG systems have to incrementally adapt to newly emerging knowledge. We record the update cost and evaluate all methods on both the base queries and the new queries. The results are summarized in Table~\ref{tab:local_result_new}. TG-RAG achieves the best performance on both base and new queries, and  maintains almost consistent performance on base queries between the original corpus and the updated corpus, with only a negligible drop compared to other methods. This indicates that our approach effectively models time-sensitive knowledge in scenarios with continuously evolving information. Moreover, in contrast to GraphRAG which incurs a high indexing cost of 30.0 million prompt and 7.8 million completion tokens due to complete graph summary regeneration, our method requires only 1.6 million prompt and 2.0 million completion tokens by updating the summaries for new time points and their corresponding ancestor nodes, resulting in a significantly lower update cost. Overall, these findings confirm that our method adapts well to evolving knowledge, preserving retrieval accuracy while efficiently integrating new information, which is a crucial capability for real-world QA systems that evolve over time.

\subsubsection{Results of Abstract Question Answering} We evaluate the performance of different methods on abstract question answering, as shown in Table~\ref{tab:global_result}. For the dimension of Temporal Coverage, our approach achieves remarkable win rates of 0.889 against GraphRAG, 0.764 against HippoRAG2, and 0.986 against LightRAG on base queries over the base corpus. This strong performance in temporal aspects directly reflects the effectiveness of our temporal-aware graph construction and retrieval framework. For Comprehensiveness and Diversity dimension, our method also shows substantial advantages with consistently high win rates, highlighting its ability to generate detailed and diverse answers.

To illustrate the differences qualitatively, Table~\ref{tab:case} presents a representative case comparing GraphRAG and our method. When answering ``Why did Iron Mountain Incorporated maintain consistent EBITDA margin growth from 2020 Q1 to 2022 Q4?'', our TG-RAG method demonstrates clear superiority in temporal reasoning by explicitly organizing the explanation along the temporal progression from 2020 to 2022. This temporal structuring enables comprehensive coverage of the entire growth narrative while maintaining logical coherence. The LLM judge ranks our answer higher across all dimensions, proving the effectiveness of our method.

\begin{table*}[t]
\caption{A representative case comparing GraphRAG to our method on abstract question answering.\label{tab:case}}
\centering
\begin{tabular}{|l|p{15cm}|}
\toprule 
\textbf{Query} & Why did Iron Mountain Incorporated maintain consistent EBITDA margin growth from 2020 Q1 to 2022 Q4? \\ \midrule
\textbf{GraphRAG} & This performance is supported by a diversified business model encompassing physical storage, records and information management (RIM), data centers, digital solutions, and asset lifecycle management (ALM). \#\#\# Key Drivers of Performance
\textbf{Data Center Business Growth:} IRM's Global Data Center Business has been a significant contributor to its financial success. This segment has experienced robust growth, with strong leasing activity and increased demand for bandwidth, partly driven by remote working trends. The adjusted EBITDA margin for the data center business has been reported at a healthy 45.8\%, with year-on-year growth reaching 21\% and 25\% in different periods. Financial strategies such as capital recycling and sale-leaseback transactions have been employed to fuel the expansion of this business.
\textbf{Productivity and Mix Improvements:} Cost productivity and a strong mix of business have been identified as key drivers for improving the Adjusted EBITDA margin across operations. These factors directly influence profitability and are central to the company's strategy for enhancing financial health...
\textbf{Challenges and Offsetting Factors:} While the company has shown improvements, certain factors have partially offset gains in the Adjusted EBITDA margin. These include fixed cost deleverage, which occurs when lower service revenue leads to less efficient utilization of fixed costs, and higher bonus compensation accrual...
 \\ \midrule
\textbf{TG-RAG} & Various factors played a vital role, including capital recycling programs, digital transformations, and innovative projects that enhanced operational efficiency. \textbf{In 2020}, Iron Mountain faced the initial pandemic challenges but simultaneously began laying the groundwork for future growth. The company emphasized cost management strategies and operational efficiency improvements. \#\#\# Key Initiatives \textbf{Cost Management}: Strategic projects that focused on reducing operational costs started to take effect. \textbf{Capital Recycling}: A strong approach to capital recycling set the stage for improved financial flexibility. \textbf{By 2021}, Iron Mountain had escalated its initiatives focusing on its data center business and digital transformation. This included the launch of innovative products like Smart Sort and Secure ITAD... \textbf{As 2022 progressed}, Iron Mountain reported notable financial performance improvements, achieving \$1.28 billion in revenue during Q4 2022... the adjusted EBITDA was positively influenced by innovations and strategic partnerships, such as those established with a global content delivery network. The continued enhancement of core services, including data center offerings, supported margin expansion with notable year-on-year growth in service revenues. \#\# \textbf{Conclusion}: From 2020 Q1 to 2022 Q4, Iron Mountain Incorporated's consistent EBITDA margin growth can be attributed to a multifaceted strategy that combined operational excellence, strategic project execution, and a robust pipeline of innovative products and customer contracts. By focusing on cost management, capital recycling, and expanding digital infrastructures, ..., resulting in a strong EBITDA growth narrative. \\ \midrule
\textbf{LLM Decision} & \textbf{Comprehensiveness}: Winner: TG-RAG; \textbf{Diversity}: Winner: TG-RAG; \textbf{Temporal Coverage}: Winner: TG-RAG; \textbf{Overall Winner}: Winner: TG-RAG\\ \bottomrule
\end{tabular}
\end{table*}

\begin{table*}[t]
\caption{Ablation study on specific question answering (base queries on the base corpus). \label{tab:ablation}}
\centering
\begin{tabular}{cccccccc}
\toprule
\multirow{2}*{Temporal Retrieval} & \multirow{2}*{PPR Ranking} & \multirow{2}*{Temporal Indexing} &\multicolumn{3}{c}{LLM Metrics}&\multicolumn{2}{c}{Non-LLM Metrics}\\
&  &  & Correct \textuparrow & Refusal & Incorrect \textdownarrow & ROUGE-L \textuparrow & $\mathrm{F}_1$ \textuparrow\\ \midrule
\checkmark & \checkmark  & \checkmark  & \textbf{0.599} & 0.191 & \textbf{0.210} & \textbf{0.493}  & \textbf{0.490} \\ \midrule
\checkmark  & \texttimes  & \checkmark & 0.580& 0.223&0.197 &0.483 &0.472 \\
\texttimes  & \checkmark  & \checkmark &0.382 &0.423 &0.195 &0.376 &0.356 \\
\texttimes  & \texttimes  & \checkmark &0.482 &0.294 &0.223 & 0.434& 0.416\\
\texttimes  & \texttimes  & \texttimes & 0.381& 0.458& 0.161& 0.359& 0.345\\
\bottomrule
\end{tabular}
\end{table*}

\begin{figure}[t!]
\centering
\includegraphics[width=0.9\linewidth]{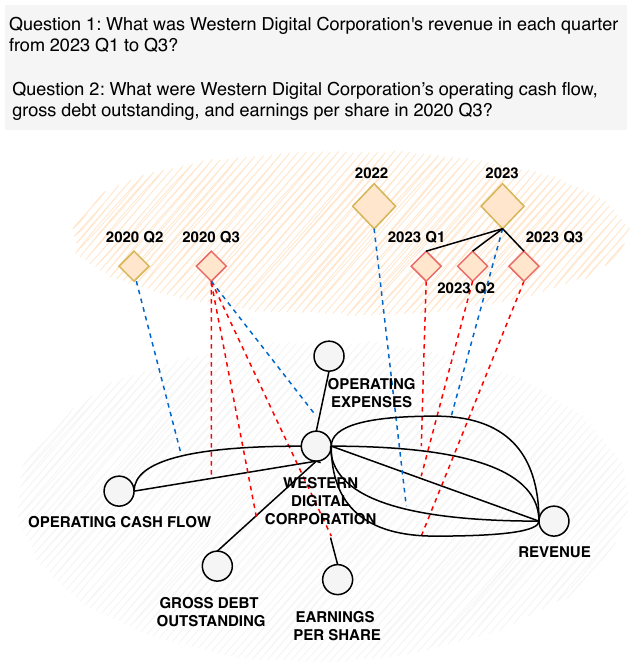}
\caption{Visualization of a subgraph sampled from our constructed bi-level temporal graph. 
Dashed lines represent the connections between upper-layer temporal nodes and lower-layer relational edges, 
where red lines indicate the temporal clues used to answer given query examples, and blue lines denote others present in the graph.}
\label{fig:case-visualization}
\end{figure}

\subsection{Why does our method perform well?} \label{exp:Q2} 
To understand the strong performance of our method, we conduct a detailed analysis from both quantitative and qualitative perspectives. First, we perform ablation studies to quantitatively evaluate the contribution of each core component in our framework. Second, we present a graph visualization with statistical comparisons to illustrate how our bi-level temporal structure effectively represents temporal facts while maintaining structural compactness.

\subsubsection{Ablation Study}
To evaluate the contribution of each core component, we conduct an ablation study on specific question answering using the base queries over the base corpus. 
Our model consists of three key modules: Temporal retrieval performs query-centric time filtering to select facts within the temporal scope $T^q$, PPR Ranking applies Personalized PageRank on the subgraph to propagate relevance scores from the seed entities and obtain more informative entity-level rankings; and Temporal Indexing, which builds time-stamped relation quadruples for precise evidence representation and retrieval. Four ablation variants are compared against the full model. 
\begin{itemize}
    \item \textit{w/o PPR Ranking}, which removes graph propagation and directly assigns edge scores using relation and query similarity. Specifically, the original edge scoring function in Eq.~(2) is replaced by 
$s(\varepsilon) = \mathbf{1}[\tau \in T^q]\,\gamma_{\varepsilon}$, for each $\varepsilon=(v_1,v_2, r, \tau)\in\mathcal{E}^q$.
    \item \textit{w/o Temporal Retrieval}, which disables query-centric time identification, running standard PageRank over all nodes to calculate entity scores and defining edge scores without temporal constraints. The original Eq.~(1) and Eq.~(2) is modified to $s(v)=\textsc{PR}\big(\mathcal{G}_K^q\big)$, $s(\varepsilon)=s(v_1)+s(v_2)$.
    \item \textit{w/o Temporal Retrieval + PPR}, which removes both time filtering and graph propagation, relying solely on relation and query similarity. The original edge scoring function in Eq.~(2) is replaced by $s(\varepsilon) = \gamma_{\varepsilon}$. 
    \item \textit{w/o All}, which further disables the temporal graph indexing module, degrading the graph construction from temporal to static. Retrieval in this setting is performed over a non-temporal graph without timestamped relations.
    \end{itemize}

The experimental results are shown in Table~\ref{tab:ablation}. First, removing only PPR Ranking results in a relatively modest performance drop compared to the full model, indicating that graph propagation helps refine local relevance. Second, when Temporal Retrieval is disabled, Correct score drops substantially from 0.599 to 0.382 and Refusal rate increases significantly to 0.423, highlighting the importance of temporal filtering for handling time-sensitive queries. Without proper time scoping, the retrieval system is overwhelmed by temporally irrelevant evidence, which subsequently misleads the answer generation process. Interestingly, the variant removing both Temporal Retrieval and PPR Ranking performs better than removing only Temporal Retrieval. This suggests that without time filtering, regular pagerank would propagate noisiness and dilute relevance, leading to poorer chunk scores and worse answers. The consistent superiority of the full model  demonstrates that the three components work synergistically: temporal retrieval ensures relevant fact positioning, PPR propagates these relevant signals effectively, and temporal indexing provides the necessary structural foundation for precise evidence representation.

\subsubsection{Graph Visualization and Statistics}
To intuitively illustrate how our method repesents and leverages temporal information, we visualize a small subgraph extracted from the constructed bi-level temporal graph, as shown in Figure~\ref{fig:case-visualization}. The detailed edge descriptions are provided in the Appendix. The upper layer (orange region) represents time nodes, while the lower layer (gray region) depicts entities and their associated relations. A key advantage of this design is that the graph can represent the same factual relation across different time periods via multiple time-stamped edges. For example, the relation between Western Digital Corporation and Revenue can appear under 2023 Q1, Q2, and Q3 as distinct temporal instances. This fine-grained temporal modeling effectively captures the factual evolution of the same knowledge across time, allowing the model to distinguish between time-specific evidence and thus prevent temporal confusion during retrieval and answer generation. For instance, given the query \textit{``What was Western Digital Corporation's revenue in each quarter from 2023~Q1 to Q3?''}, 
our model first identifies the temporal nodes 2023~Q1, 2023~Q2, and 2023~Q3, and retrieves the respective company--revenue relations, yielding accurate results. 
Similarly, for the query \textit{``What were Western Digital Corporation’s operating cash flow, gross debt outstanding, and earnings per share in 2020~Q3?''}, 
our model first identifies the temporal node 2020~Q3 and successfully locates the correct facts. These visualized examples clearly demonstrate that the proposed temporal graph structure enables effective time-aware retrieval and precise retrieval for time sensitive question answering.

\begin{table}[t]
\caption{Comparison of graph sizes constructed by different methods, measured by the number of entities and relations extracted from the base corpus and the updated corpus.\label{tab:graph_statistics}}
\centering
\begin{tabular}{lcccc}
\toprule
\multirow{2}*{Model}&\multicolumn{2}{c}{Base Corpus}&\multicolumn{2}{c}{Updated Corpus}\\
& \#entity & \#relation & \#entity & \#relation\\ \midrule
GraphRAG & 45,540 & 59,679 & 54,854 & 71,989 \\
LightRAG &35,003 & 42,892 & 42,514 &53,383 \\
HippoRAG2 &19,318 & 24,205& 22,128&28,153 \\
Ours &16,817 & 28,157 & 20,110 & 34,671 \\
\bottomrule
\end{tabular}
\end{table}

As shown in Table~\ref{tab:graph_statistics}, our method encodes the same corpus with substantially fewer entities and relations, resulting in lower storage costs and faster retrieval. This compactness stems from our temporal design, which shares entity nodes across time while introducing timestamped connections to capture evolution, thus effectively eliminating structural redundancy.

\subsection{Applicability to Alternative LLMs and Generic QA Datasets} \label{exp:Q3}
To further examine the generalization of our approach, we apply it to different LLMs and extend the evaluation to generic QA datasets. 
Table~\ref{tab:diff_llms} reports the performance of various QA LLMs on base queries over the base corpus. 
Our method maintains strong performance across diverse model architectures, notably achieving a 0.625 Correct score with DeepSeek-R1. It is worth noting that even when using the open-source Llama-3.3-70B-Instruct, our approach still outperforms other baselines built upon GPT-4o-mini. The effectiveness across both proprietary and open-source models indicates that our approach does not critically depend on specific LLM implementations, making it flexible and accessible for different practical deployments.

\begin{table}[t]\renewcommand\tabcolsep{0.88pt}
\caption{Evaluation of different LLMs for query answering: base queries on the base corpus. \label{tab:diff_llms}} % graph construction and
\centering
\begin{tabular}{lccccc}
\toprule
QA LLM & Correct & Refusal & Incorrect & ROUGE-L & $\mathrm{F}_1$\\
\midrule
GPT-4o-mini  & 0.599 & 0.191 & 0.210 & 0.493  & 0.490 \\
%GPT-4o &0.591 &0.210 & 0.199 & 0.483 & 0.481 \\
Llama-3.3-70B-Instruct  &0.466 & 0.373 & 0.160 & 0.419 & 0.403 \\
Qwen3-235B-A22B-Instruct-2507 &0.619&0.200&0.180&0.500&0.482\\
DeepSeek-R1  &  0.625& 0.174 & 0.201 & 0.493 &0.468  \\ 
\bottomrule
\end{tabular}
\end{table}

\begin{table}[t]
\centering
\caption{QA Performance ($\mathrm{F}_1$ scores) across Different RAG Datasets using Llama-3.3-70B-Instruct as the Index and QA LLMs. NV-Embed-v2 are adopted as the embedding model, following the setting in \cite{gutierrez2025from}. Results marked with $\dag$ are reported by \cite{gutierrez2025from}. \label{tab:more_rag_datasets}}
\begin{tabular}{lccc}
\toprule
Model & NaturalQuestions & 2WikiMultihopQA & HotpotQA \\
\midrule
NaiveRAG$\dag$ &0.619 & 0.615 & 0.753 \\
GraphRAG$\dag$ & 0.469 & 0.586 & 0.686 \\
LightRAG$\dag$ &0.166 & 0.116 & 0.024 \\
HippoRAG2$\dag$ &0.633 & 0.710 & 0.755  \\
Ours & \textbf{0.648} & \textbf{0.719} & \textbf{0.757}\\
\bottomrule
\end{tabular}
\end{table}

In addition, we evaluate the general applicability of our approach on three widely-used RAG datasets, NaturalQuestions \cite{DBLP:journals/tacl/KwiatkowskiPRCP19}, 2WikiMultihopQA  \cite{DBLP:conf/coling/HoNSA20}, and HotpotQA  \cite{DBLP:conf/emnlp/Yang0ZBCSM18}. To ensure a consistent and fair comparison, we follow the configuration of HippoRAG2 \cite{gutierrez2025from} by using Llama-3.3-70B-Instruct for both indexing and question answering, and NV-Embed-v2 \cite{DBLP:conf/iclr/Lee0XRSCP25} as the embedding model. For graph indexing, we prompt the model to extract not only temporal quadruples but also non-temporal factual triples. During retrieval, we remove temporal filtering when the query lacks a clearly identifiable time scope, enabling our method to operate robustly across generic QA tasks. As shown in Table~\ref{tab:more_rag_datasets}, our method achieves $\mathrm{F}_1$ scores of 0.648 on NaturalQuestions, 0.719 on 2WikiMultihopQA, and 0.757 on HotpotQA, confirming its broad applicability and robustness beyond time-sensitive QA.

\section{Conclusion}
In this paper, we identify and address a critical yet often overlooked challenge in retrieval-augmented generation: managing the temporal dynamics of real-world knowledge. We propose Temporal GraphRAG (TG-RAG), a novel framework that represents external corpora as bi-level temporal graphs, where timestamped relations capture factual evolution and hierarchical time summaries encode multi-granular trends. The framework incorporates time-aware retrieval strategies that dynamically select relevant temporal contexts to enhance answer accuracy. Furthermore, we contribute a new benchmark ECT-QA and an incremental evaluation protocol that simulates realistic corpus growth, allowing assessment of retrieval accuracy and update efficiency. Extensive experiments demonstrate the effectiveness and efficiency of our method. We believe this work provides a concrete step toward making RAG systems temporally aware and practically deployable in dynamic environments.

\section*{Acknowledgments} The authors used a generative AI tool for language polishing. Specifically, OpenAI’s ChatGPT\footnote{\url{https://chatgpt.com/}} was used to improve grammar and clarity in the \emph{Abstract} and \emph{Introduction}. All text was reviewed and verified by the authors.

\bibliographystyle{IEEEtran}
\bibliography{refs}

\clearpage
\appendices
\section*{Temporal Extraction Prompt}

\begin{center}
%\fcolorbox{black}{gray!10}{\parbox{1\linewidth}{
%}

\begin{lstlisting}[language=Python]
-Goal-
Given a user query that is potentially ask a time-related question, identify the timestamp entities in the query, follows the structure:
- entity_name: standard format of the timestamp entity identified in context, following {timestamp_format}
- entity_type: {timestamp_types}
- temporal_logic: one of <at, before, after, between>
If temporal_logic is <between>, format a pair of timestamp entities as ("entity"{tuple_delimiter}<between>{tuple_delimiter}<entity_name>{tuple_delimiter}<entity_name>{tuple_delimiter}<entity_type>)
If temporal_logic is <at, before, after>, format a pair of timestamp entities as ("entity"{tuple_delimiter}<temporal_logic>{tuple_delimiter}<entity_name>{tuple_delimiter}<entity_type>)

When finished, output {completion_delimiter}

######################
-Examples-
######################
Example 1:
User Query:
Who was the CEO of DXC Technology on January 1, 2022?
################
Output:
("entity"{tuple_delimiter}"at"{tuple_delimiter}"2022-01-01"{tuple_delimiter}"date"){completion_delimiter}
#############################
Example 2:
User Query:
What strategic decisions were made between Q2 and Q4 2022?
#############
Output:
("entity"{tuple_delimiter}"between"{tuple_delimiter}"2022-Q2"{tuple_delimiter}"2022-Q4"{tuple_delimiter}"quarter"){completion_delimiter}

#############################
Example 3:
User Query:
What has changed in Aon's leadership after the NFP acquisition in 2023?
#############
Output:
("entity"{tuple_delimiter}"after"{tuple_delimiter}"2022"{tuple_delimiter}"year"){completion_delimiter}
#############################

Output:
-Real Data-
######################
User Query: {input_text}
######################
Output:
\end{lstlisting}

%}
\end{center}

\section*{Local Query Prompt}
\begin{center}
\begin{lstlisting}[language = Python]
---Role---
You are a helpful assistant responding to questions about temporal data in the relevant chunks provided.

---Goal---
Answer the user's question based on the available information in the given chunks. Your response should:

1. **Answer the question directly** - Provide the specific information requested
2. **Be temporally accurate** - Ensure temporal information matches the question's scope
3. **Be evidence-based** - Only provide answers when you have clear, unambiguous evidence

**Critical Guidelines:**
- **ONLY answer if you have clear, unambiguous evidence** from the provided chunks
- **Refuse to answer** if the evidence is unclear, conflicting, incomplete, or uncertain
- **Refuse to answer** if you cannot find specific information requested in the question
- **Refuse to answer** if the chunks contain related information but not the exact answer needed
- For temporal queries, be flexible with temporal expressions (e.g., "2023 Q4" vs "fourth quarter of 2023")
- Use the given chunks as your primary source of information
- The chunks are ordered by relevance, so focus on the most relevant chunks

**When to refuse (respond with "No explicit evidence for the question"):**
- No relevant information found in chunks
- Information is present but unclear or ambiguous
- Information is incomplete for the specific question asked
- You are uncertain about the answer
- Cannot find the specific information requested

---Chunks--- (Ordered by relevance)

{"".join(processed_chunks)}

---Query---
{query}

---Target response length and format---

{response_format}
\end{lstlisting}
\end{center}
\section*{Global Query Prompt}
\begin{center}
\begin{lstlisting}[language=Python]
---Role---
You are a helpful assistant responding to questions about a dataset by synthesizing perspectives from multiple analysts.


---Goal---
Generate a response of the target length and format that responds to the user's question, summarize all the reports from multiple analysts who focused on different parts of the dataset.

Note that the analysts' reports provided below are ranked in the **descending order of importance**.

If you don't know the answer or if the provided reports do not contain sufficient information to provide an answer, just say so. Do not make anything up.

The final response should remove all irrelevant information from the analysts' reports and merge the cleaned information into a comprehensive answer that provides explanations of all the key points and implications appropriate for the response length and format.

Add sections and commentary to the response as appropriate for the length and format. Style the response in markdown.

The response shall preserve the original meaning and use of modal verbs such as "shall", "may" or "will".

Do not include information where the supporting evidence for it is not provided.

Create a comprehensive analysis that demonstrates diversity through:

**STRUCTURAL DIVERSITY:**
- Use varied section headers (mix of descriptive, analytical, and thematic headers)
- Alternate between different organizational patterns within the same response
- Combine narrative flow with analytical rigor
- Mix chronological and thematic organization

**PRESENTATION DIVERSITY:**
- Vary paragraph styles (some analytical, some narrative, some data-focused)
- Use different evidence presentation methods (direct quotes, summaries, bullet points, tables)
- Mix formal analysis with accessible explanations
- Alternate between macro and micro perspectives

**CONTENT DIVERSITY:**
- Present multiple analytical frameworks and perspectives
- Include both quantitative and qualitative insights
- Balance strategic, operational, and financial viewpoints
- Show both short-term and long-term implications

**STYLISTIC DIVERSITY:**
- Vary sentence structures and lengths
- Mix technical precision with engaging narrative
- Use different transition styles between sections
- Combine data-driven analysis with strategic insights

---Target response length and format---
{response_type}


---Analyst Reports---
{report_data}

---Target response length and format---
{response_type}

Add sections and commentary to the response as appropriate for the length and format. Style the response in markdown.    
\end{lstlisting}
\end{center}

\section*{Local Query Prediction Evaluation Prompt}
\begin{center}
\begin{lstlisting}[language = Python]
You are a fact-level evaluation assistant.

Given:
- A user's question
- The ground-truth answer (unanswerable)
- The model's prediction (which may contain multiple factual claims)

Task:
1. Identify all **distinct factual elements** in the prediction answer (e.g., numbers, facts, etc.).
2. For each element, compare with the model's prediction and classify as one of:
   - Correctly Refusal: The model explicitly refused to answer this element (e.g., said "I don't know" or "No explicit evidence").
   - Incorrect: The model provided a wrong or hallucinated value.

Output:
- Count how many elements are correctly refusal or incorrect.
- Use exact matching unless it's obvious the meaning is equivalent.
- Only consider content explicitly present in the prediction. Do not infer missing values.
        """

    USR_TEMPLATE = """
### Input
**Question:**  
{question}

**Ground-Truth Answer:**  
{answer}

**Model Prediction:**  
{prediction}

### Assistant
Return the result in **strict JSON format** as:
{{
  "correctly refusal": <int>,
  "incorrect": <int>
}}
\end{lstlisting}
\end{center}

\section*{Global Query Prediction Evaluation Prompt}
\begin{center}
\begin{lstlisting}[language = Python]
---Role---
You are an expert tasked with evaluating two answers to the same question based on three criteria: **Comprehensiveness**, **Diversity**, and **Temporal Coverage**.

---Goal---
You will evaluate two answers to the same question based on three criteria: **Comprehensiveness**, **Diversity**, and **Temporal Coverage**.

- **Comprehensiveness**: How much detail does the answer provide to cover all aspects and details of the question?

- **Diversity**: Evaluate the richness and variety of the answer across multiple dimensions:
   - **Content Diversity**: Does the answer explore different aspects, angles, or facets of the question rather than repeating similar points?
   - **Analytical Perspectives**: Does it present multiple viewpoints, analytical frameworks, or methodological approaches?
   - **Evidence Variety**: Does it draw from diverse sources, data types, or evidence categories?
   - **Structural Variety**: Does the presentation style, organization, or formatting vary appropriately to the content?
   - **Depth vs. Breadth Balance**: Does it strike a good balance between detailed analysis and broad coverage?
   - **Innovation**: Does it offer unique insights or creative approaches to addressing the question?
   Note: Length alone does not indicate diversity. Focus on the richness of perspectives and approaches.

- **Temporal Coverage**: How well does the answer capture the **time dimension** of the question?
   - Does it clearly reference the relevant periods (years, quarters, events) mentioned or implied in the query?
   - Are chronological relationships accurate and complete?
   - Is the timeline explanation easy to follow and logically organized?

For each criterion, choose the better answer (either Answer 1 or Answer 2) and explain why. Then, select an overall winner based on these three categories.

Here is the question:
{question}

Here are the two answers:

**Answer 1:**
{prediction1}

**Answer 2:**
{prediction2}

Evaluate both answers using the three criteria listed above and provide detailed explanations for each criterion.

Output your evaluation in the following JSON format:

{{
    "Comprehensiveness": {{
        "Winner": "[Answer 1 or Answer 2]",
        "Explanation": "[Provide explanation here]"
    }},
    "Diversity": {{
        "Winner": "[Answer 1 or Answer 2]",
        "Explanation": "[Provide explanation here]"
    }},
    "Temporal Coverage": {{
        "Winner": "[Answer 1 or Answer 2]",
        "Explanation": "[Provide explanation here]"
    }},
    "Overall Winner": {{
        "Winner": "[Answer 1 or Answer 2]",
        "Explanation": "[Summarize why this answer is the overall winner based on the three criteria]"
    }}
}}
\end{lstlisting}
\end{center}

\section*{Case Study of Temporal Graph Visualization}
\subsection*{Example of Timestamped Relations}
Table~\ref{tab:case_edges} lists the detailed edge descriptions for the visualization example shown in Figure~\ref{fig:case-visualization}. 
Each record corresponds to a temporal quadruple $(v_1, v_2, r, \tau)$ representing a factual relation grounded at a specific time. This example is constructed from the subgraph of \textit{Western Digital Corporation} and illustrates how our bi-level temporal graph organizes multi-temporal financial facts.
\begin{table*}[t!]
\caption{Timestamped relations extracted for the Western Digital Corporation case study.\label{tab:case_edges}}
\centering
\renewcommand{\arraystretch}{1.05}
\begin{tabular}{p{3.5cm}p{3.5cm}p{1.5cm}p{6.5cm}}
\toprule
\textbf{Entity 1} & \textbf{Entity 2} & \textbf{Time} & \textbf{Relation Description} \\ 
\midrule
Western Digital Corporation & Operating Cash Flow & 2020~Q2 & In 2020~Q2, Western Digital Corporation's operating cash flow was \$257~million. \\
Western Digital Corporation & Operating Cash Flow & 2020~Q3 & In Q3 2020, Western Digital Corporation's operating cash flow was \$142 million. \\
Western Digital Corporation & Gross Debt Outstanding & 2020~Q3 & At the end of Q3 2020, Western Digital Corporation's gross debt outstanding was \$9.8 billion. \\
Western Digital Corporation & Earnings Per Share & 2020~Q3 & In Q3 2020, Western Digital Corporation's earnings per share was \$0.85. \\
Western Digital Corporation & Operating Expenses & 2020~Q3 & In Q3 2020, Western Digital Corporation's operating expenses were \$738 million, slightly lower than expected.\\
Western Digital Corporation & Revenue & 2022 & For the full fiscal year 2022, Western Digital Corporation's revenue was \$18.8 billion, an 11\% increase from fiscal 2021. \\
Western Digital Corporation & Revenue & 2023 & For fiscal year 2024, Western Digital's total revenue was \$13 billion, up 6\% from fiscal year 2023. \\
Western Digital Corporation & Revenue & 2023~Q1 & In Q1 2023, Western Digital Corporation reported total revenue of \$3.7 billion, down 17\% sequentially and 26\% year over year. \\
Western Digital Corporation & Revenue & 2023~Q2 & In fiscal Q2 2023, Western Digital Corporation delivered revenue of \$3.1 billion, which was at the high end of its guidance range.\\
Western Digital Corporation & Revenue & 2023~Q3 & In Q3 2023, Western Digital Corporation reported total revenue of \$2.8 billion, which was down 10\% sequentially and 36\% year over year. \\
\bottomrule
\end{tabular}
\end{table*}

\subsection*{Case Study Description}

\begin{table*}[t]
\caption{Case study examples and evaluation metrics on ECT-QA.\label{tab:case_study}}
\centering
\footnotesize
\setlength{\tabcolsep}{3pt}
\renewcommand{\arraystretch}{1.2}
\begin{tabularx}{\linewidth}{@{}X X X c c c c c@{}}
\toprule
\textbf{Question} & \textbf{Golden Answer} & \textbf{Prediction} & \textbf{Correct} & \textbf{Refusal} & \textbf{Incorrect} & \textbf{ROUGE-L} & $\mathbf{F}_{\mathbf{1}}$ \\
\midrule
What were Western Digital Corporation’s operating cash flow, gross debt outstanding, and earnings per share in 2020~Q3? &
\$142~million, \$9.8~billion, and \$0.85. &
 Operating cash flow: \$142 million, gross debt outstanding: \$9.8 billion, earnings per share: \$0.85.
 &1.0   &0.0   & 0.0  & 0.583&0.500\\
What was Western Digital Corporation's revenue in each quarter from 2023 Q1 to Q3? &
\$3.7~billion, \$3.1~billion, and \$2.8~billion. &2023 Q1: \$3.7 billion, 2023 Q2: \$3.1 billion, 2023 Q3: \$2.8 billion.
 &1.0
 &0.0   &0.0   &0.720   &0.632  \\
\bottomrule
\end{tabularx}
\end{table*}

Table~\ref{tab:case_study} details two representative question answering cases from the ECT-QA benchmark. 
Each case demonstrates how our temporal graph structure supports accurate retrieval and reasoning over time-dependent financial facts. 
\subsubsection*{Single-time Specific Query}
Consider the query: \textit{``What were Western Digital Corporation’s operating cash flow, gross debt outstanding, and earnings per share in 2020 Q3?''}

Our retrieval process is as follows:
\begin{enumerate}
    \item Query-Centric Time Identification: The query is analyzed to identify the relevant time point \textit{2020-Q3}.
    \item Dynamic Subgraph Positioning: TG-RAG retrieves top-30 relevant relations in the temporal knowledge graph and identifies temporally focused seed set anchored to the \textit{2020-Q3} time node.
    \item Local Retrieval: Top 5 nodes with highest PPR score are ``WESTERN DIGITAL CORPORATION", ``2020-Q3",``GROSS MARGIN", ``REVENUE", ``FREE CASH FLOW", after chunk scoring the chunk with highest score comes from the earnings call transcript from company Western Digital Corporation in 2020-Q3.
\end{enumerate}
The retrieved evidences make the QA LLM to generate the accurate answer: \textit{``Operating cash flow: \$142 million, gross debt outstanding: \$9.8 billion, earnings per share: \$0.85.''}
\\
\subsubsection*{Multi-Time Query}
Now consider a  query spanning multiple time points: \textit{``What was Western Digital Corporation's revenue in each quarter from 2023 Q1 to Q3?''}

Our method handles this as follows:
\begin{enumerate}
    \item Query-Centric Time Identification: The query is recognized as requiring data from three distinct time points: \textit{2023-Q1}, \textit{2023-Q2}, and \textit{2023-Q3}.
    \item Dynamic Subgraph Positioning: TG-RAG retrieves top-20 relevant relations in the temporal knowledge graph and identifies temporally focused seed set anchored to the time nodes.
    \item Local Retrieval: Top 5 nodes with highest PPR score are ``REVENUE'', ``WESTERN DIGITAL CORPORATION'',``2023-Q1'',``2023-Q3'', after chunk scoring the top 3 chunks with highest scores come from the earnings call transcripts from company Western Digital Corporation in 2023-Q1, 2023-Q2 and 2023-Q3 respectively.
\end{enumerate}

The retrieved evidences enable the LLM to generate the accurate answer: \textit{``\$142 million, \$9.8 billion, and \$0.85.''}\\
\\
This case study confirms that our graph structure is not merely a storage format but an effective retrieval backbone. By explicitly modeling time as a first-class citizen, it prevents temporal confusion and provides the necessary granularity to answer complex, time-sensitive questions precisely.

\end{document}